\documentclass[aps,twocolumn, noshowpacs, floatfix]{revtex4}
\usepackage{}

\usepackage{amsfonts}
\usepackage{amssymb}
\usepackage{graphicx}
\usepackage{amsmath}
\usepackage[english]{babel}
\usepackage{color}

\begin{document}

\title{Effect of off-diagonal exciton-phonon coupling on intramolecular singlet fission}

\author{Zhongkai Huang, Yuta Fujihashi, and Yang Zhao\footnote{Electronic address:~\url{YZhao@ntu.edu.sg}}}
\affiliation{Division of Materials Science, Nanyang Technological University, Singapore 639798, Singapore}

\begin{abstract}
Intramolecular singlet fission (iSF) materials provide remarkable advantages in terms of tunable electronic structures, and quantum chemistry studies have indicated strong electronic coupling modulation by high frequency phonon modes. In this work, we formulate a microscopic model of iSF with simultaneous diagonal and off-diagonal coupling to high-frequency modes. A non-perturbative treatment, the Dirac-Frenkel time-dependent variational approach is adopted using the multiple Davydov trial states. It is shown that both diagonal and off-diagonal coupling can aid efficient singlet fission if excitonic coupling is weak, and fission is only facilitated by diagonal coupling if excitonic coupling is strong. In the presence of off-diagonal coupling, it is found that high frequency modes create additional fission channels for rapid iSF. Results presented here may help provide guiding principles for design of efficient singlet fission materials by directly tuning singlet-triplet interstate coupling.
\end{abstract}
\date{\today}

\maketitle

\section{Introduction}

Singlet fission (SF) is a multielectron process in which a singlet exciton generated by light irradiation is converted to two triplet excitons.\cite{smith_2010, smith_2013, chan_2013}
In 1965, SF was first coined to explain photophysics in anthracene crystals.\cite{singh_1965} In recent years, interest in SF has been renewed because of its potential to increase maximum efficiency of organic solar cells.\cite{hanna_2006, congreve2013external, piland_2014} As a result, SF has been studied in various organic materials of polyacenes,\cite{ramanan_2011} polyenes,\cite{billon_2013} and other chromophores, such as perylenediimide and tert-butyl-substituted terrylenes.\cite{eaton_2013, eaton_2015}

To date, most efforts have been dedicated to understanding intermolecular singlet fission (xSF), in which a singlet state on one molecule couples with the ground-state of neighboring molecules to form an intermolecular correlated triplet pair.
The xSF mainly involves conventional SF materials, such as crystalline solids of pentacene,\cite{wilson2011ultrafast,chan2011observing} tetracene,\cite{chan2012energy,wilson2013temperature} and other organic materials.\cite{ma2012rubrene,eaton_2013}
Mechanisms of xSF have been the focus of many ultrafast spectroscopic measurements\cite{chan_2013, wilson2011ultrafast, chan2011observing, chan2012energy, wilson2013temperature} and extensive theoretical studies based on dynamics simualtions\cite{greyson_2010, teichen2012microscopic, berkelbach_2013_1, berkelbach_2013_2, berkelbach2014microscopic3, mirjani2014theoretical, tao2014electronically, tamura2015first, fujihashi2016fluctuations, yao2016coherent} and electronic structures calculations.\cite{zimmerman2010singlet, zimmerman_2011, feng2013fission, yost2014transferable, casanova2014electronic} However, due to intermolecular nature of xSF, the efficiency of this process is highly sensitive to geometric stacking, crystal environment, side-group, and other factors.\cite{renaud2013mapping, wang2014maximizing,tamura2015first, lukman_2016_sidegroup}
Devices based on xSF which manipulate crystal packing are limited by the lack of high throughput processing strategies of developing highly ordered molecular structures.
Difficulties in engineering molecular packing morphology have promoted the development of intramolecular singlet fission (iSF), in which the two long-lived triplets are located on the same molecule.\cite{monahan2015charge} Achieved in $2015$,\cite{busby2015intraSF} iSF materials offer great advantages in terms of tunable molecular and electronic structures, and have included a series of chromophore dimers with a conjugated linker,\cite{ethan_2014, sanders2016intraSF, varnavski2015intraSF, tao_2016, tao_2016_1, tao_2016_1} such as covalently coupled pentacenes,\cite{eric_2016, samuel2016intraSF} and a covalent tetracene dimer.\cite{ethan_2014} Recent transient absorption measurements for diphenyl-dicyano-oligoene groups (DPDC$_{n}$) molecules have shown that xSF occurs in DPDC$_{n}$ in acetonitrile solution, while in DPDC$_{n}$ solid films, iSF dominates.\cite{meisner2012dpdc, tuan_2014_crystaleffect} Based on these findings, Trinh {\it et al.} suggested that efficient SF can be achieved by independent tuning of singlet-triplet pair coupling and triplet pair splitting.\cite{tuan_2014_crystaleffect} However, a limited understanding of detailed xSF and iSF mechanisms hinders the design of versatile SF materials. In particular, a unified treatment of phonon effects remains elusive \cite{berkelbach_2013_1, berkelbach_2013_2, yuta_2017, eric_2016, tao_2016}.

In organic crystals, fluctuations in electronic energies are induced by intramolecular vibrations.\cite{berkelbach_2013_1,berkelbach_2013_2} (Note that this type of exciton-phonon coupling is often called diagonal coupling.)
Recently, ultrafast spectroscopic measurements in the xSF materials have shown that phonon modes coupled to electronic excitations play a crucial role in the xSF process.\cite{musser_2015, bakulin2016real, monahan2017dynamics}
In particular, high-frequency modes of pentacene derivatives\cite{yuta_2017, tempelaar2017vibronic1, tempelaar2017vibronic2} and crystalline tetracene \cite{morrison_2017, elenewski2017functional} are found to facilitate efficient fission by resonances between vibrational modes and energy splittings of electronic states. On the other hand, intermolecular vibrations of the crystals induce off-diagonal exciton-phonon coupling which modulates the electronic coupling between the singlet and triplet pair state. Berckelbach {\it et. al} have considered the off-diagonal coupling in acene crystals and demonstrated that it plays a minor role because frequencies of the intermolecular vibrations are significantly lower than the energy difference between the singlet and the triplet pair state.\cite{berkelbach_2013_2}
Effects of those forms of exciton-phonon coupling on xSF are usually restricted to particular materials such as perlenediimide crystals\cite{renaud_2014}.
In contrast, in the context of iSF, the transition between the singlet state and triplet pair state occurs within a covalently linked dimer, and thus intramolecular vibrations of the linker part may induce fluctuations in the electronic coupling.
Indeed, quantum chemical calculations of the covalent tetracene dimer demonstrated that high frequency intramolecular vibrations induce nonnegligible off-diagonal coupling as well as diagonal coupling.\cite{ethan_2014}
The two kinds of coupling have been found to be tunable in a typical iSF molecule by changing linker types and by engineering dihedral angles between the chromophore units and the linker.\cite{eric_2016, samuel2016intraSF, sanders2016intraSF}
The corresponding iSF dynamics has been obtained by treating the exciton states quantum mechanically and phonons classically, indicating that SF time scales vary with the linker types.\cite{tao_2016_1, tao_2016}
Those investigations on exciton-phonon coupling are believed to have helped understand fast iSF observed in a broad range of organic molecules.
However, detailed iSF mechanisms under the influence of simultaneous diagonal and off-diagonal exciton-phonon coupling remain ill understood, and thus a full quantum dynamical investigation is required for the elucidation of this issue.

Impacts of off-diagonal coupling on exciton dynamics in organic crystals have been investigated previously by the Munn-Silbey theory \cite{mu_85} and a variational method using the Davydov D$_2$ {\it Ansatz} \cite{zh_12,zh_97} Recently, Zhao and coworkers have developed a refined trial state, the multiple Davydov D$_2$ {\it Ansatz}, to accurately treat dynamics of the generalized Holstein model with simultaneous diagonal and off-diagonal coupling.\cite{zhou2015polaron, zhou2016fast} Within the framework of the Dirac-Frenkel time-dependent variation, accuracy of the method can be carefully monitored by quantifying how faithfully our result follows the time-dependent Schr\"odinger equation.
In this work, this variational method will be employed to explore effects of off-diagonal coupling on iSF dynamics, and to demonstrate that high frequency phonon modes may open up additional fission channels for rapid iSF in the presence of off-diagonal coupling.

In this letter, we focus on a dimer model of the iSF dynamics on the basis of the four-electron four-orbital basis.\cite{monahan2015charge,smith_2010}
A simple scheme is considered for the iSF process, $\left| \rm g\right\rangle \rightarrow\left| \rm S_{1}\right\rangle \rightarrow\left| {\rm TT}\right\rangle$, where $\left| \rm g\right\rangle$ denotes the electronic ground state, $\left|{\rm S}_{1}\right\rangle$ is the singlet state, and $\left| {\rm TT}\right\rangle$ represents the correlated triplet pair state.
In some organic materials, the iSF and xSF processes may be accelerated by a mediated pathway, in which the singlet state converts to a triplet pair state via the charge transfer ($\rm CT$) state.
Quantum chemistry calculations of acene derivatives have demonstrated that the energy of the $\rm CT$ state is significantly higher than that of the singlet excited states,\cite{ zimmerman_2011,casanova2014electronic} and $\rm CT$ states have been found not to participate in xSF process as actual intermediates between $\rm S_1$ and $\rm TT$.\cite{berkelbach_2013_2,mirjani2014theoretical,tao2014electronically,tamura2015first, fujihashi2016fluctuations,yao2016coherent} Moreover, using electronic structures calculations and transient absorption spectroscopy, iSF has been demonstrated to occur via a direct coupling mechanism that is independent of the $\rm CT$ states in the covalent pentacene dimer\cite{eric_2016} and tetracene dimer.\cite{ethan_2014}
Therefore, we assume that the sole effect of the $\rm CT$ states is to effectively couple the $\rm S_1$ and $\rm TT$ states.
We employ a system-bath Hamiltonian describing the both diagonal and off-diagonal coupling,
\begin{align}
\hat{H}=\hat{H}_{\rm sys}+\hat{H}_{\rm bath}+\hat{H}_{\rm sys-bath}.
\label{Hamiltonian}
\end{align}
First term of $\hat{H}$ is the system Hamiltonian, and is chosen to be that of an electronically diabatic Hamiltonian for
$\left| {\rm g} \right\rangle$, $\left| {\rm S}_{1}\right\rangle$ and $\left| {\rm TT}\right\rangle$
\begin{align}
\hat{H}_{\rm sys}=\sum_{n={\rm S_1, TT}} \epsilon_{n{\rm g}} | n \rangle  \langle n |
+ \sum_{m={\rm S_1, TT}} \sum_{n\neq m}   J_{mn} | m \rangle  \langle  n | ,
\label{Hamiltonian_S}
\end{align}
where $\epsilon_{n{\rm g}}$ is the Franck-Condon energy associated with electronic transition from $ | \rm g \rangle $ to $ | n \rangle $, and $J_{\rm S_1, TT}$ is strength of the interstate coupling between ${\rm S}_1$ and ${\rm TT}$.
$J_{\rm S_1, TT}$ includes the contribution of the direct coupling between ${\rm S}_1$ and ${\rm TT}$ based on the two electron integrals\cite{smith_2010,berkelbach_2013_1} as well as that of the effective coupling created by quantum mixing of ${\rm CT}$ and electronic states.
Second term of $\hat{H}$ represents the bath Hamiltonian $\hat{H}_{\rm bath}$, and is given by
\begin{align}
\hat{H}_{\rm bath}=\sum_{q}\hbar\omega_{q}\hat{b}_{q}^{\dagger}\hat{b}_{q},
\end{align}
where $\omega_q$ indicates the frequency of the $q$-th mode of the bath with creation operator, $\hat{b}_{q}^{\dagger}$, and annihilation operator, $\hat{b}_{q}$.
Third term of $\hat{H}$ represents system-bath coupling, $\hat{H}_{\rm sys-bath}$, and is given by
\begin{align}
\hat{H}_{\rm sys-bath}=&\sum_{n={\rm S_1, TT}} | n \rangle  \langle n | \sqrt{\lambda_{n,{\rm g}}}
\cdot  \hat{\mathcal{E}}_x \notag \\
&+  \sum_{m={\rm S_1, TT}}  \sum_{n\neq m}    | m \rangle  \langle  n |  \sqrt{\lambda_{\rm S_1, TT}^{\rm  o.d.}}
\cdot \hat{\mathcal{E}}_y,
\end{align}
where we have defined operators $\hat{\mathcal{E}}_x =  \hbar\omega_{q} g_{q}(\hat{b}_{q}^{\dagger}+\hat{b}_{q})$ and $\hat{\mathcal{E}}_y =  \hbar\omega_{q} c_{q}(\hat{b}_{q}^{\dagger}+\hat{b}_{q})$.
$g_{q}$ and $c_{q}$ are the diagonal and off-diagonal exciton-phonon coupling strength between the system and $q$-th mode, respectively.
$\lambda_{mn}$ represents the reorganization energy associated with the transition from $| m \rangle$ to $| n \rangle$, and $\lambda_{mn}^{\rm  o.d.}$ is the amplitude of fluctuations in interstate coupling between $| m \rangle$ and $| n \rangle$.
Details of both types of coupling are given in Appendix.
The diagonal coupling describes fluctuations in the electronic energies induced by intramolecular vibrations, whereas the term of the off-diagonal coupling attributes to the fluctuations in electronic coupling induced by intramolecular and intermolecular vibrations, as mentioned above.
The spectral density $J_\alpha(\omega)$ $(\alpha=x,y)$ is a useful measure for characterizing various forms of exciton-phonon coupling, and can be evaluated in terms of $g_{q}$ ($c_{q}$) as
\begin{align}
J_{x}(\omega )=\frac{\pi}{2}\sum_{q}\hbar \omega_{q}^{2} g_{q}^{2}\delta (\omega-\omega_{q} ),
\end{align}
\begin{align}
J_{y}(\omega )=\frac{\pi}{2}\sum_{q} \hbar \omega_{q}^{2} c_{q}^{2}\delta (\omega-\omega_{q} ).
\end{align}
In this study, we model the diagonal coupling  spectral densities using underdamped Brownian oscillators with Huang-Rhys factor, $S_{m}=\lambda_{m{\rm g}} / (\hbar\omega_{\rm diag})$, such that
\begin{align}
J_{x}(\omega )=
\frac{4 \gamma_{\rm diag} \omega_{\rm diag}^2 \omega }{ (\omega^{2} - \omega_{\rm diag}^{2} )^2  +4 \gamma_{\rm diag}^2 \omega^2}
 ,
\end{align}
where $\omega_{\rm diag}$ is the vibrational frequency and $\gamma_{\rm diag}$ is the vibrational relaxation rate.
In the iSF materials, the dominant contributions to off-diagonal coupling are from the intramolecular vibrations.
Similarly, we model the off-diagonal coupling spectral densities as
\begin{align}
J_{y}(\omega )=
\frac{4 \gamma_{\rm o.d.} \omega_{\rm o.d.}^2 \omega }{ (\omega^{2} - \omega_{\rm o.d.}^{2} )^2  +4 \gamma_{\rm o.d.}^2 \omega^2}
 ,
\end{align}
where $\omega_{\rm o.d.}$ is the vibrational frequency and $\gamma_{\rm o.d.}$ is the vibrational relaxation rate.

The multiple Davydov trial states with multiplicity $M$, which are essentially $M$ copies of the corresponding single Davydov {\it Ansatz},\cite{bera2014stabilizing,zhou2014ground,zhou2015polaron,zhou2016fast,wang2016variational} have been developed for the Holstein model and the spin-boson model.
In the two level system description of the iSF process, one of the multiple Davydov trial states, the multi-${\rm D}_2$ {\it Ansatz} with multiplicity $M$, can be constructed as\cite{zhou2015polaron,zhou2016fast,huang2017polaron,huang2017}
\begin{align}\label{D2_state}
\left|{\rm D}_{2}^{M}\right\rangle =\sum_{i=1}^{M} \sum_{n={\rm S_1, TT}} c_{in}(t)\left| n \right\rangle e^{(\sum_{q} f_{iq}(t)\hat{b}_{q}^{\dagger}-{\rm H.c.})} \left|0\right\rangle_{\rm vib} 
\end{align}
where ${\rm H.c.}$ denotes the Hermitian conjugate, and $\left|0\right\rangle_{\rm vib}$ is the vacuum state of the bosonic bath. $c_{in}(t)$ is time-dependent variational parameter for the amplitudes in states $\left| n \right\rangle$, and $f_{iq}(t)$ denotes the phonon displacements, where $i$ and $q$ label the $i$-th coherent superposition state and $q$-th effective bath mode, respectively. If $M=1$, the multi-${\rm D}_2$ {\it Ansatz}
is reduced to the usual Davydov ${\rm D}_2$ trial state.
The time dependent variational parameters $c_{in}(t)$ and $f_{iq}(t)$ are determined by adopting the Dirac-Frenkel variational principle.
Detailed derivation of equations of motion for variational parameters are given in references elsewhere \cite{zhou2015polaron,zhou2016fast,wang2016variational,huang2017polaron} and the Appendix.

To obtain numerical solutions to the equations of motion for the variational parameters, the continuum spectral densities $J_x(\omega)$ and $J_y(\omega)$ need to be discretized.
In this study, the method of linear discretization is employed.
The displacement $g_q$ for each $\omega_q$ is given by $g_{q}^{2}=2J_{x}\left(\omega_{q}\right)\Delta\omega/(\pi\hbar\omega_{q}^{2})$.
With respect to off-diagonal coupling, the displacement $c_q$ for each $\omega_q$ is given by $c_{q}^{2}=2J_{y}\left(\omega_{q}\right)\Delta\omega/(\pi\hbar\omega_{q}^{2})$.
The validity of our variational method for SF dynamics is carefully examined by quantifying how faithfully our result follows the Schr\"odinger equation in balance with the computational efficiency, as details are given in the Appendix.

Below we present and discuss numerical results regarding iSF dynamics in the dimer model. A simple dimer model including two excitonic states $\rm S_{1}$ and $\rm TT$ is adopted, and our focus is on the effect of two intramolecular vibration modes, one of which is diagonally coupled ($\hbar\omega_{\rm diag}$) to the exciton states, and the other, off-diagonally coupled ($\hbar\omega_{\rm o.d.}$). A Huang-Rhys factor of $0.7$ is chosen for $\rm S_{1}$, which is estimated from fitting measured absorption spectra of acene derivatives with a theoretical spectroscopic model.\cite{yamagata2011, beljonne2013, chen2016optical} It is found that the reorganization energy of $\rm TT$ is several times larger than that of $\rm S_{1}$ in pentacene derivatives and tetracenes, and the off-diagonal coupling strength is one order of magnitude smaller than the diagonal coupling strength,\cite{tamura2015first, ito_2015} such that the Huang-Rhys factor of $\rm TT$ is set at $S_{\rm TT}=2S_{\rm S_{1}}$ throughout this work, and the off-diagonal coupling Huang-Rhys factor $S_{\rm S_1,TT}^{\rm o.d.}=\lambda_{\rm S_1, TT}^{\rm o.d.}/(\hbar\omega_{\rm o.d.})$ is chosen to be $0.1$.
To be in line with the beating lifetime due to vibrational coherence in $2$D electronic spectra of pentacene derivatives,\cite{bakulin2016real} the vibrational relaxation rates are set to $\gamma^{-1}_{\rm diag}=\gamma^{-1}_{\rm o.d.}=1~{\rm ps}$. As the initial condition for our numerics, only the singlet state is excited according to the Franck-Condon principle.
It has been suggested that efficient SF can be achieved by tuning the $\rm S_{1}$-$\rm TT$ interstate coupling and triplet pair splitting independently.\cite{tuan_2014_crystaleffect} As shown in our Hamiltonian~(\ref{Hamiltonian_S}), excitonic coupling $J_{\rm S_1, TT}$ determines the direct $\rm S_{1}$-$\rm TT$ coupling. If the frequencies of the vibrational modes $\omega_{\rm diag}$ and $\omega_{\rm o.d.}$ are high compared with the thermal energy $k_{\rm B}T$, the intramolecular vibrations are thermally inactivated, and the fission dynamics driven by the high frequency modes is temperature independent in a wide temperature range.\cite{yuta_2017} Thus, in this study, temperature is set to be $T=0$ to reduce the numerical cost, although the inclusion of the temperature effect in the multiple Davydov {\it Ansatz} is straightforward by applying Monte Carlo importance sampling.\cite{wang2016variational}

\begin{figure}[tbp]
\centering
\includegraphics[scale=0.47]{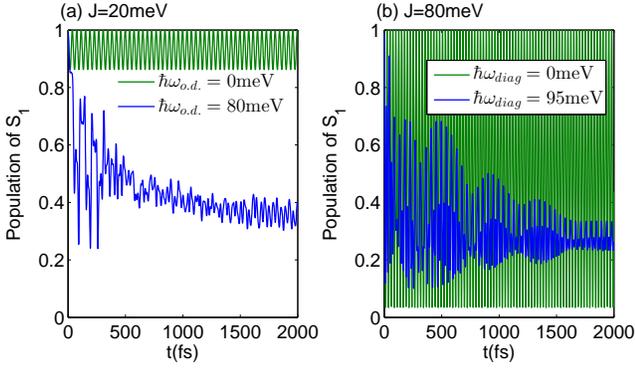}
\caption[FIG]{Time evolution of the singlet population for (a) the case of weak excitonic coupling $J_{\rm S_1,TT}=20~{\rm meV}$, $\epsilon_{\rm S_{1},g}-\epsilon_{\rm TT,g}=100~{\rm meV}$, $\hbar\omega_{\rm diag}=0~{\rm meV}$, and $\hbar\omega_{\rm o.d.}=80~{\rm meV}$, and (b) the case of strong excitonic coupling $J_{\rm S_1,TT}=80~{\rm meV}$, $\epsilon_{\rm S_{1},g}-\epsilon_{\rm TT,g}=30~{\rm meV}$, $\hbar\omega_{\rm diag}=95~{\rm meV}$, and $\hbar\omega_{\rm o.d.}=0~{\rm meV}$. The green lines correspond to cases in the absence of exciton-phonon coupling.}
\label{population_J}
\end{figure}

We first look into a scenario with weak excitonic coupling $J_{\rm S_1,TT}=20~{\rm meV}$ and the transition energy $\epsilon_{\rm S_{1},g}-\epsilon_{\rm TT,g}=100~{\rm meV}$, both of which are typical values for the SF process in pentacene derivatives.\cite{berkelbach_2013_2, bakulin2016real, tamura2015first} As shown in Fig.~\ref{population_J}(a), the oscillation amplitude of population of $\rm S_{1}$ is $0.13$ in the absence of exciton-phonon coupling (green line). Despite weak direct coupling, it is suggested that strong mixing between $\rm S_{1}$ and $\rm TT$ can be allowed with the assistance of exciton-phonon coupling.\cite{berkelbach_2013_1, eric_2016} In the presence of diagonal exciton-phonon coupling the high frequency phonon modes have been shown to facilitate the efficient SF if excitonic coupling is weak.\cite{yuta_2017} Here we study the impact of off-diagonal coupling on SF dynamics. As shown in Fig.~\ref{population_J}(a), the population of $S_{1}$ for $\hbar\omega_{\rm o.d.}=80~{\rm meV}$, calculated by the ${\rm D}_2^{M=3}$ {\it Ansatz}, decays to $0.4$ at long times, signaling an efficient SF process. For this scenario, the emission of a single off-diagonally coupled phonon can relax the initial singlet excitation to a double triplet state.\cite{renaud_2014} Consequently, the presence of off-diagonal coupling changes the SF process substantially if excitonic coupling is weak.

Next we turn to a case with strong excitonic coupling $J_{\rm S_1,TT}=80~{\rm meV}$ and transition energy $\epsilon_{\rm S_{1},g}-\epsilon_{\rm TT,g}=30~{\rm meV}$, as strong excitonic coupling has been achieved experimentally via interchromophore bridge control in covalently linked tetracene dimers.\cite{vallet2013tunable}
As shown in Fig.~\ref{population_J}(b), in the absence of exciton-phonon coupling, the $\rm S_{1}$ population exhibits purely Rabi oscillations (green line), due to  strong excitonic coupling. If only diagonal exciton-phonon coupling is added, the picture is changed drastically (blue line), and the system becomes trapped in the $\rm TT$ state. This rapid, irreversible decay of the Rabi oscillation in the SF process is owing to dissipation induced by diagonal exciton-phonon coupling, in agreement with decay dynamics in the ultrafast charge transfer process at an oligothiophene-fullerene heterojunction.\cite{tamura2011chargetransfer, tamura2012chargetransfer} However, efficient SF is not found under the influence of off-diagonal exciton-phonon coupling if excitonic coupling is strong (See the Appendix).

\begin{figure}[tbp]
\centering
\includegraphics[scale=0.5]{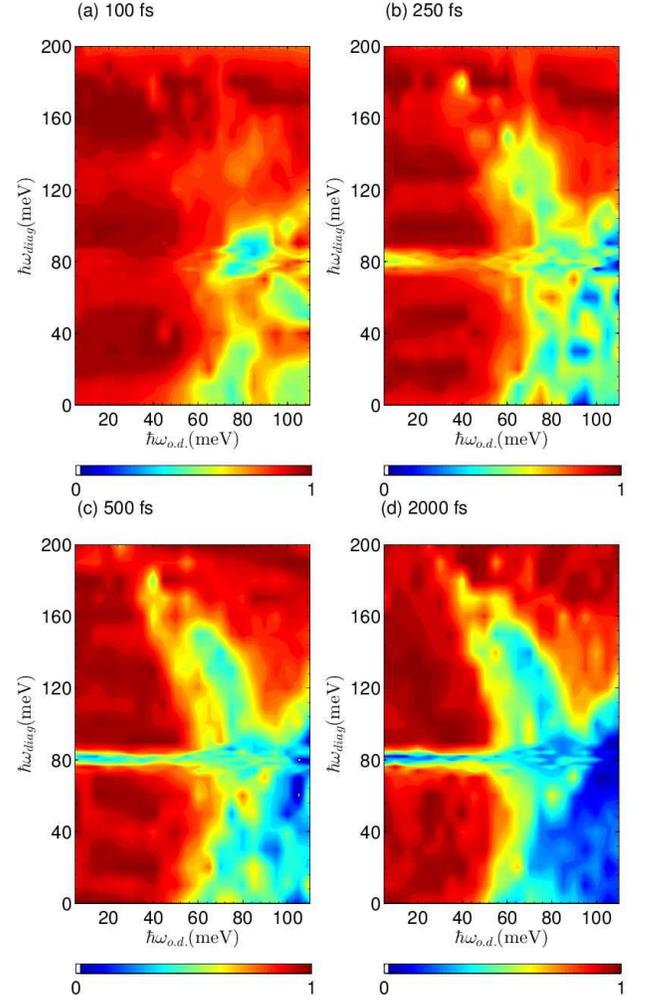}
\caption[FIG]{Snapshots of singlet population as functions of $\hbar\omega_{\rm diag}$ and $\hbar\omega_{\rm o.d.}$ at the time of (a) $100~{\rm fs}$, (b) $250~{\rm fs}$, (c) $500~{\rm fs}$ and (d) $2000~{\rm fs}$. The other parameters are $J_{\rm S_1,TT}=20~{\rm meV}$ and $\epsilon_{\rm S_{1},g}-\epsilon_{\rm TT,g}=100 ~{\rm meV}$.}
\label{population0_190_5_80}
\end{figure}

In particular, for recently developed iSF materials, such as covalent pentacene and diazadiborine dimers,\cite{eric_2016, tao_2016, ethan_2014} excitonic coupling strength is often smaller than the transition energy, so we choose $J_{\rm S_1,TT}=20~{\rm meV}$ and $\epsilon_{\rm S_{1},g}-\epsilon_{\rm TT,g}=100~{\rm meV}$ in our iSF dynamics study. Off-diagonal coupling has been found for low-frequency phonon modes in teracene,\cite{ito_2015, renaud_2014, tamura2015first} and high-frequency phonon modes in covalent chromophore dimers.\cite{ethan_2014, eric_2016, tao_2016}
In order to further explore effects of off-diagonal coupling on iSF, we examine time evolution of singlet population as functions of $\hbar\omega_{\rm diag}$ and $\hbar\omega_{\rm o.d.}$. Fig.~\ref{population0_190_5_80} shows snapshots of singlet population at $t = 100$, $250$, $500$, and $2000~{\rm fs}$. On one hand, a single phonon mode of $\hbar\omega_{\rm diag}=80~{\rm meV}$ brings about a single efficient channel for SF dynamics in the absence of off-diagonal coupling.\cite{yuta_2017} On the other hand, Fig.~\ref{population0_190_5_80} clearly exhibits some channels for rapid SF dynamics due to the presence of off-diagonal coupling, despite a complex dependence of efficient SF on $\hbar\omega_{\rm o.d.}$. In order to better interpret this dependence, we divide the phase space into three regions, $5\leq\hbar\omega_{\rm o.d.}\leq50~{\rm meV}$, $50 < \hbar\omega_{\rm o.d.}\leq80~{\rm meV}$, and $80<\hbar\omega_{\rm o.d.}\leq110~{\rm meV}$.

\begin{figure}[tbp]
\centering
\includegraphics[scale=0.47]{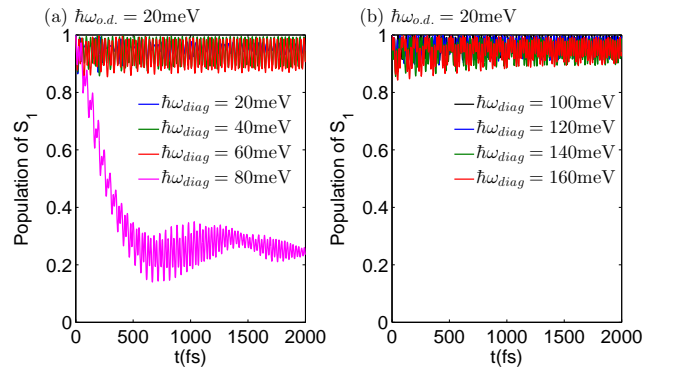}
\caption[FIG]{Time evolution of singlet population for the the phonon mode $\hbar\omega_{\rm diag}=20~{\rm meV}$ diagonally coupled to the $\rm S_{1}$ and $\rm TT$ states. The phonon modes off-diagonally coupled are (a)$\hbar\omega_{\rm o.d.}=20, 40, 60$ and $80~{\rm meV}$, (b) $\hbar\omega_{\rm o.d.}=100, 120, 140$ and $160~{\rm meV}$. The other parameters are same as Fig.~\ref{population0_190_5_80}}
\label{hw1=20}
\end{figure}

If $5\leq\hbar\omega_{\rm o.d.}\leq50~{\rm meV}$, the SF dynamics in Figs.~\ref{population0_190_5_80}(c) and \ref{population0_190_5_80}(d) is driven by the phonon modes around $\hbar\omega_{\rm diag}=80~{\rm meV}$.
Fig.~\ref{hw1=20} shows the time evolution of singlet population for $\hbar\omega_{\rm o.d.}=20~{\rm meV}$ and four values of $\hbar\omega_{\rm diag}$. The dependence of SF on $\hbar\omega_{\rm diag}$ here is qualitatively similar to that in the only diagonal coupling scenario.\cite{yuta_2017} Due to the low frequency phonon mode $\hbar\omega_{\rm o.d.}$, the value of $\lambda_{\rm S_1,TT}^{\rm o.d.}=S_{\rm S_1,TT}^{\rm o.d.}\hbar\omega_{\rm o.d.}$ is significantly lower compared with the diagonal reorganization energy, and thus SF is dominated by the resonances between vibrational modes $\hbar\omega_{\rm diag}$ and electronic splittings.

\begin{figure}[tbp]
\centering
\includegraphics[scale=0.47]{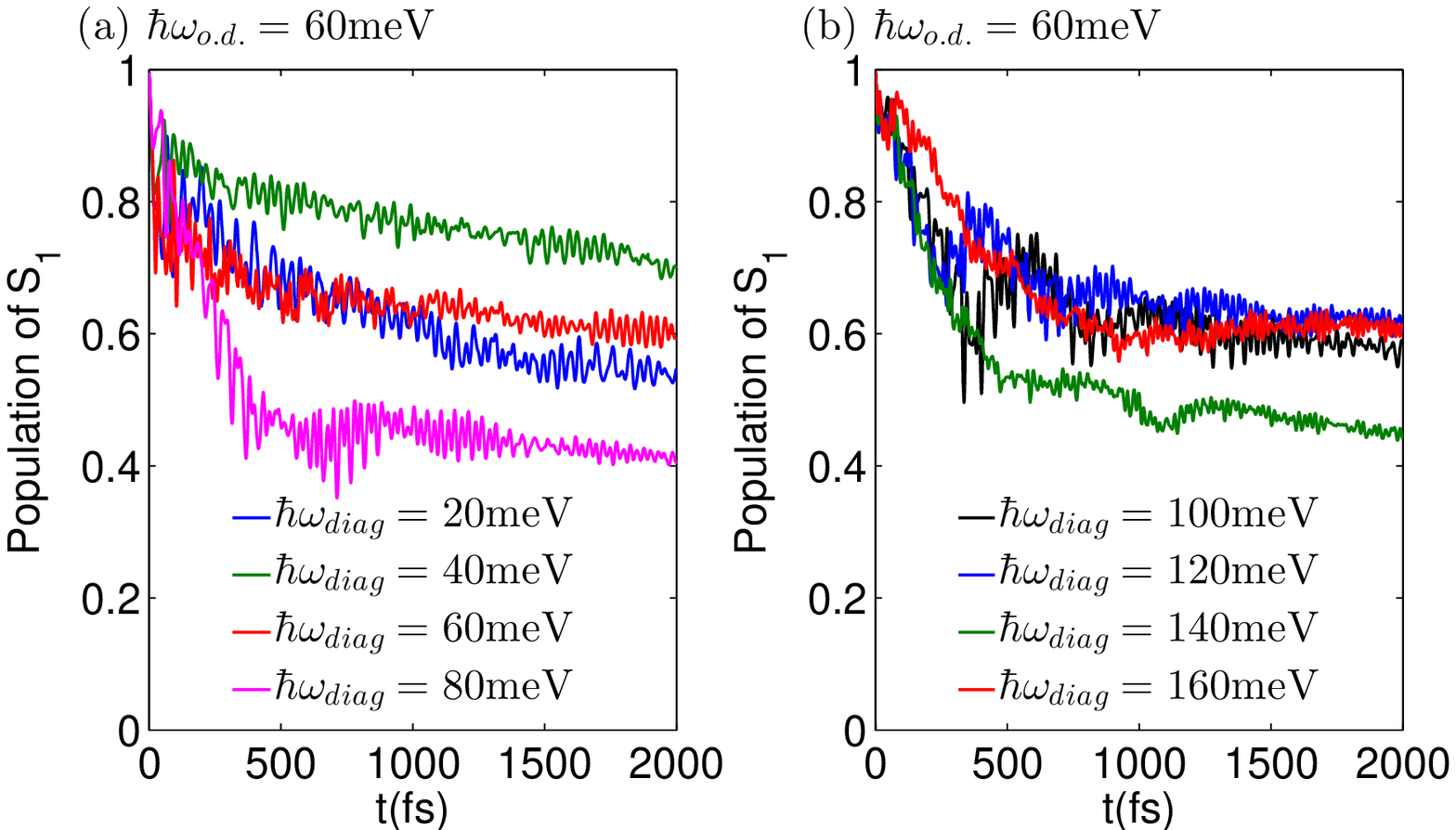}
\caption[FIG]{Time evolution of singlet population for the the phonon mode $\hbar\omega_{\rm diag}=60~{\rm meV}$ diagonally coupled to the $\rm S_{1}$ and $\rm TT$ states. The phonon modes off-diagonally coupled are (a)$\hbar\omega_{\rm o.d.}=20, 40, 60$ and $80~{\rm meV}$, (b) $\hbar\omega_{\rm o.d.}=100, 120, 140$ and $160~{\rm meV}$. The other parameters are same as Fig.~\ref{population0_190_5_80}}
\label{hw1=60}
\end{figure}

Concerning the intermediate region of $50 < \hbar\omega_{\rm o.d.}\leq80~{\rm meV}$, efficient SF is found in Fig.~\ref{population0_190_5_80} via several fission channels, including that around $\hbar\omega_{\rm diag}=80~{\rm meV}$. Fig.~\ref{hw1=60} shows the time evolution of singlet population for $\hbar\omega_{\rm o.d.}=60~{\rm meV}$ and various values of $\hbar\omega_{\rm diag}$. The magenta line of $\hbar\omega_{\rm diag}=80~{\rm meV}$ in Fig.~\ref{hw1=60}(a) and the green line of $\hbar\omega_{\rm diag}=140~{\rm meV}$ in Fig.~\ref{hw1=60}(b) exhibit fast decay dynamics. However, SF dynamics for both cases is much slower than that in Fig.~\ref{hw1=20}(a), because the interplay between fission dynamics and the mode $\hbar\omega_{\rm diag}=80~{\rm meV}$ is drastically affected by the introduction of the mode $\hbar\omega_{\rm o.d.}=60~{\rm meV}$. What is shown in Fig.~\ref{hw1=60} is compatible with the SF mechanism with two diagonal coupled phonon modes,\cite{yuta_2017} in which inclusion of a second phonon mode creates a new SF channel, and if there is only diagonal coupling, shifts the optimal phonon frequency that promotes SF dynamics. However, in the presence of off-diagonal coupling, the SF channel around $\hbar\omega_{\rm diag}=80~{\rm meV}$ is found unchanged.

\begin{figure}[tbp]
\centering
\includegraphics[scale=0.47]{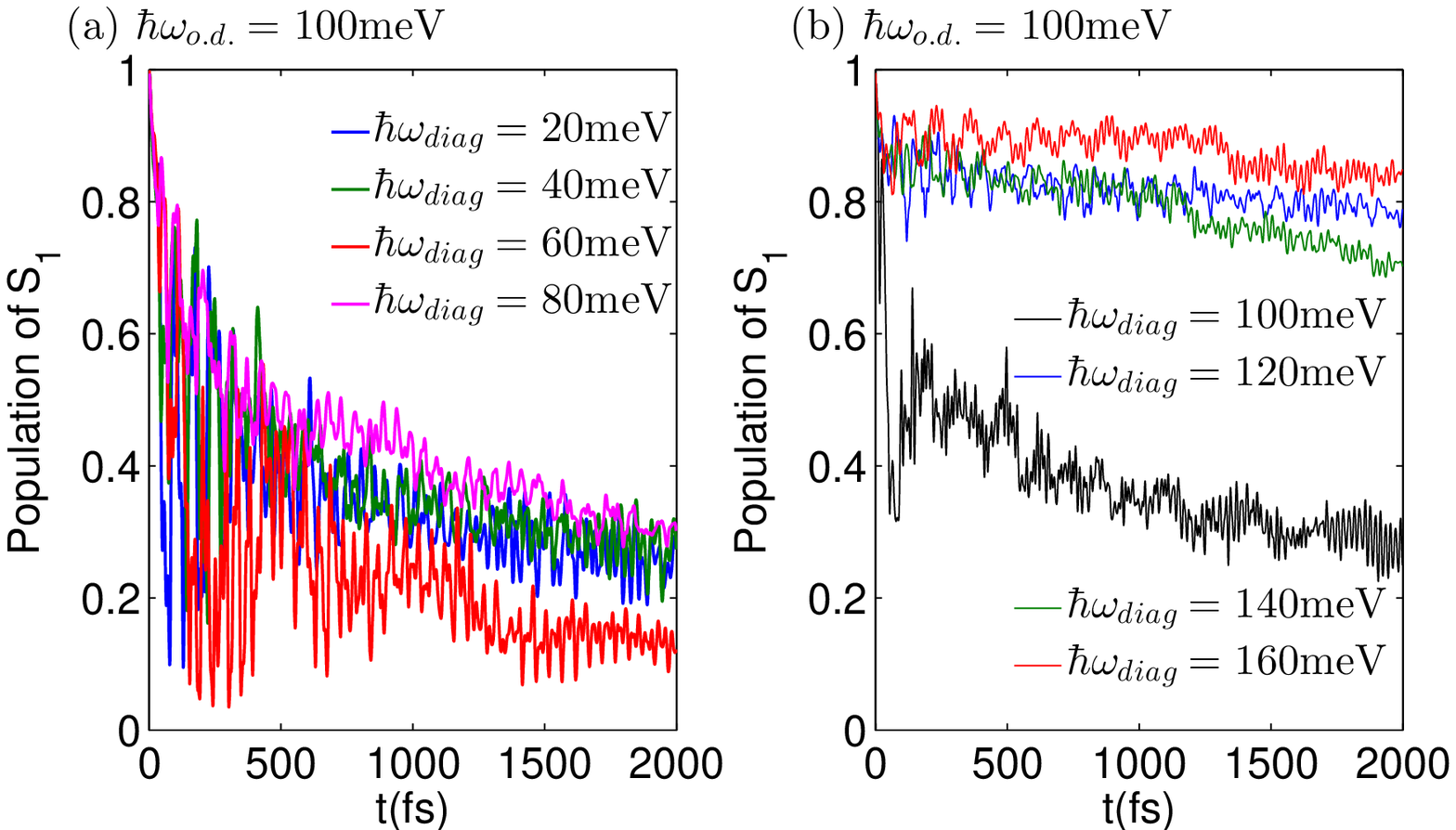}
\caption[FIG]{Time evolution of singlet population for the the phonon mode $\hbar\omega_{\rm diag}=60~{\rm meV}$ diagonally coupled to the $\rm S_{1}$ and $\rm TT$ states. The phonon modes off-diagonally coupled are (a)$\hbar\omega_{\rm o.d.}=20, 40, 60$ and $80~{\rm meV}$, (b) $\hbar\omega_{\rm o.d.}=100, 120, 140$ and $160~{\rm meV}$. The other parameters are same as Fig.~\ref{population0_190_5_80}}
\label{hw1=100}
\end{figure}

As for the third region, SF dynamics is mainly facilitated by the high frequency phonon mode ($\hbar\omega_{\rm o.d.}$), as shown in Fig.~\ref{population0_190_5_80}. Comparing to the first and second region, efficient SF in this region is found to be dependent on a larger number of phonon modes, in qualitative agreement with the dependence of SF on multiple phonon modes in covalent tetracene dimers.\cite{ethan_2014} Fig.~\ref{hw1=100} presents the time evolution of singlet population for $\hbar\omega_{\rm o.d.}=100~{\rm meV}$ and different values of $\hbar\omega_{\rm diag}$. Due to the high frequency phonon mode $\hbar\omega_{\rm o.d.}$, the initial oscillation amplitude is much larger than $0.13$, which is the oscillation amplitude in the absence of exciton-phonon coupling. The envelope of fast oscillations is found to decrease gradually owing to dissipation induced by exciton-phonon coupling. Moreover, the SF time increases with increasing $\hbar\omega_{\rm diag}$.

In conclusion, we have developed a model of iSF dynamics including simultaneous diagonal and off-diagonal exciton-phonon coupling. Following the Dirac-Frenkel time-dependent variational principle, we provide an accurate description for iSF using the multi-$\rm D_2$ {\it Ansatz}, a superposition of the usual Davydov $\rm D_2$ states. To our knowledge, diagonal coupling has been reported to aid efficient SF if excitonic coupling is weak,\cite{berkelbach_2013_1, yuta_2017} and it has also been shown in this work to facilitate efficient fission if excitonic coupling is strong. Furthermore, we have demonstrated for the first time that off-diagonal coupling plays a crucial role in the fission process only if excitonic coupling is weak.
It is determined that iSF dynamics strongly depends on the frequency of phonon modes in the presence of off-diagonal coupling, and high-frequency phonon modes result in efficient iSF even if the off-diagonal coupling strength is weak. Multiple SF channels can be created by the simultaneous presence of diagonal and off-diagonal coupling.
Thus a unified framework has been provided to capture the effects of diagonal and off-diagonal coupling on SF dynamics. Results presented here may help provide guiding principles for design of efficient singlet fission materials by directly tuning singlet-triplet interstate coupling.

\section*{Acknowledgments}
Support from the Singapore National Research Foundation through the Competitive Research Programme (CRP) under Project No.~NRF-CRP5-2009-04 is gratefully acknowledged.

\appendix
\section{Modeling intramolecular vibration}

The electronically diabatic Hamiltonian for the electronic ground state, the singlet state, and the correlated triplet pair state is given by
\begin{align}
\hat{H}=\sum_{n={\rm g, S_1, TT}}
 \hat{H}_{n} (\mbox{\boldmath $x$})
   | n \rangle  \langle n |
   + \sum_{m={\rm S_1, TT}}
   \sum_{n\neq m}
   \hat{U}_{mn}(\mbox{\boldmath $x$})
    | m \rangle  \langle  n |,
      \label{eq30}
 \end{align}
where $ \hat{H}_{n} (\mbox{\boldmath $x$})$ denotes the diabatic Hamiltonian for the vibrational degrees of freedom (DOFs), $\mbox{\boldmath $x$}$, when the electronic system is in state $ | n \rangle$.
$ \hat{U}_{mn}(\mbox{\boldmath $x$})$ represents the interstate coupling between $\rm S_1$ and $\rm TT$.
The excitonic coupling, $ \hat{U}_{mn}(\mbox{\boldmath $x$})$, is modulated by vibrational DOFs, and the vibrational dependence of $ \hat{U}_{mn}(\mbox{\boldmath $x$})$ gives the off-diagonal exciton-phonon coupling, as shown below.

The electronic energy of each diabatic state experiences modulations by the intramolecular vibrational motion.
Fluctuations in the electronic energies of the diabatic states are quantified by collective energy gap coordinates as
\begin{align}
\hat{u}_{mn} = \hat{H}_m(x) - \hat{H}_n(x) - \langle \hat{H}_m(x) - \hat{H}_n(x)  \rangle_n , \label{eq12}
\end{align}
where the canonical average, $\langle \hat{O} \rangle_n = {\rm Tr}\{ \hat{O} \hat{\rho}_n^{\rm eq} \} $ with $\hat{\rho}_n^{\rm eq}= e^{-\beta \hat{H}_n}/ {\rm Tr}\{e^{-\beta \hat{H}_n} \} $, has been introduced for any operator $\hat{O}$.
Here $\beta$ is the inverse temperature, $1/k_{\rm B}T$.
The reorganization energy associated with the transition from $|m\rangle$ to $|n\rangle$ can be given as
\begin{align}
\lambda_{mn} =\langle \hat{H}_m(\mbox{\boldmath $x$}) - \hat{H}_n(x)  \rangle_n - (\epsilon_m^\circ -\epsilon_n^\circ), \label{eq13}
\end{align}
where $\epsilon_m^\circ$ denotes the equilibrium energy of the diabatic state $| m \rangle$, namely $\epsilon_m^\circ=\langle \hat{H}_m(x)   \rangle_m $.

For photo-induced SF processes, the reaction coordinate associated with the initial photoexcitation, $\hat{u}_{\rm S_1,g}$, and the coordinate for the subsequent SF, $\hat{u}_{\rm TT, S_1}$, are required.
Although the equations may involve statistically orthogonal components of fluctuations, both are described by the same component of fluctuations (one-dimensional reaction coordinate model),
\begin{align}
\hat{u}_{\rm S_1,g} = - \sqrt{\lambda_{\rm S_1,g}} \cdot \hat{\mathcal{E}}_x,  \label{eq14}
\end{align}
\begin{align}
\hat{u}_{\rm TT,S_1} = -\eta_{\rm TT,S_1} \sqrt{\lambda_{\rm TT,S_1}} \cdot \hat{\mathcal{E}}_x
\label{eq7}
\end{align}
with $\eta_{\rm TT, S_1}= \cos\theta_{\rm TT,S_1}=\pm 1$.
$\eta_{\rm TT, S_1}=1$ ($-1$) corresponds to positive (negative) correlation between $\rm S_1$ and $\rm TT$, and
 $\hat{\mathcal{E}}_x$ is normalized as
\begin{align}
\langle \hat{\mathcal{E}}_x; \hat{\mathcal{E}}_x \rangle_{\rm g}= 2/\beta ,
\end{align}
where $\langle \hat{\mathcal{O}}_1; \hat{\mathcal{O}}_2 \rangle_{\rm g}$ denotes the canonical correlation of two operators $\hat{\mathcal{O}}_1$ and $\hat{\mathcal{O}}_2$, and $\langle \hat{\mathcal{O}}_1; \hat{\mathcal{O}}_2 \rangle_{\rm g}
= \beta^{-1} \int_0^\beta d\eta \langle e^{\eta H_g} \hat{\mathcal{O}}_1 e^{-\eta H_g} \hat{\mathcal{O}}_2 \rangle_{\rm g}$.
Here we assume that the vibrational DOFs, $\mbox{\boldmath $x$}$, can be modeled as a set of harmonic oscillators, and the reaction coordinates $ \hat{\mathcal{E}}_x$ have Gaussian fluctuations.
Thus, the reorganization dynamics can be characterized by the relaxation function in term of $ \hat{\mathcal{E}}_x $,
\begin{align}
\Psi_x(t) = \beta \langle \hat{\mathcal{E}}_x (0) ; \hat{\mathcal{E}}_x (t) \rangle_{\rm g} ,  \label{eq1}
\end{align}
where $\hat{\mathcal{E}}_x (t) =e^{ i H_{\rm g}t /\hbar} \hat{\mathcal{E}}_x  e^{- i H_{\rm g}t /\hbar}$.
In this work, we consider a single intramolecular vibration in molecular dimer, with frequency, $ \omega_{\rm diag}$, and Huang-Rhys factor, $S_{m,1}=\lambda_{m{\rm g }} / (\hbar\omega_{\rm diag})$.
The functional form of the relaxation function is modeled as the underdamped Brownian oscillator model,
\begin{align}
\Psi_x(t) = 2 \left( \cos \tilde{\omega}_{\rm diag}t + \frac{\gamma_{\rm diag}}{\tilde{\omega}_{\rm diag} }\sin \tilde{\omega}_{\rm diag}t  \right) e^{-\gamma_{\rm diag}t},
\end{align}
where $\gamma_{\rm diag}$ is the vibrational relaxation rate, and we have introduced $\tilde{\omega}_{\rm diag}=( \omega_{\rm diag}^2 - \gamma_{\rm diag}^2  )^{1/2}$.

In the same fashion as in the diagonal fluctuations, we consider the fluctuations in interstate coupling between electronic states.
Flutucations caused by the vibrational motion are described by the collective coordinate as
\begin{align}
\hat{u}_{mn}^{\rm o.d.} = \hat{U}_{mn}(x) - J_{mn},
\end{align}
where $ J_{mn}= \langle \hat{U}_{mn}(x) \rangle_{\rm g} $ denotes the coupling constant of averaged interstate coupling between two electronic states.
For a unified treatment, we defined the normalized reaction coordinate, $\hat{\mathcal{E}}_y$, as follows
\begin{align}
\hat{u}_{\rm S_1, TT}^{\rm o.d.} = \sqrt{\lambda_{mn}^{\rm o.d.}} \cdot \hat{\mathcal{E}}_y ,
\end{align}
where $\langle \hat{\mathcal{E}}_y; \hat{\mathcal{E}}_y \rangle_{\rm g}= 2/\beta $, and $\lambda_{mn}^{\rm o.d.}$ is the amplitude of vibrationally induced fluctuations in interstate coupling between electronic states. Huang-Rhys factor for off-diagonal coupling is $S_{mn}^{\rm o.d.}=\lambda_{mn}^{\rm o.d.}/ (\hbar\omega_{\rm o.d.})$.
The dynamical process  can be characterized by the relaxation function in term of $ \hat{\mathcal{E}}_y $,
\begin{align}
\Psi_y(t) = \beta \langle \hat{\mathcal{E}}_y (0) ; \hat{\mathcal{E}}_y(t) \rangle_{\rm g} , \label{eq2}
\end{align}
where $\hat{\mathcal{E}}_y (t) =e^{ i H_{\rm g}t /\hbar} \hat{\mathcal{E}}_y  e^{- i H_{\rm g}t /\hbar}$.
The functional form of the relaxation function is modeled as the underdamped Brownian oscillator model,
\begin{align}
\Psi_y(t) = 2 \left( \cos \tilde{\omega}_{\rm o.d.}t + \frac{\gamma_{\rm o.d.}}{\tilde{\omega}_{\rm o.d.} }\sin \tilde{\omega}_{\rm o.d.}t  \right) e^{-\gamma_{\rm o.d.}t},
\end{align}
where $\gamma_{\rm vib2}$ is the vibrational relaxation rate, and $\tilde{\omega}_{\rm o.d.}=( \omega_{\rm o.d.}^2 - \gamma_{\rm o.d.}^2  )^{1/2}$ has been introduced.
Substituting Eqs.~(\ref{eq12}), (\ref{eq13}), (\ref{eq14}) and (\ref{eq7}) into Eq.~(\ref{eq30}), the total Hamiltonian of the above subsection is recast as Eq.~(1) in the main text.

\setcounter{equation}{0}

\section{Validity of variational method}
\label{relative_deviation_section}
To quantify the accuracy of the variational dynamics based on the multiple Davydov trial states, we introduce a deviation vector $\vec{\delta}(t)$ defined as
\begin{eqnarray}
\vec{\delta}(t) & =  & \vec{\chi}(t) -\vec{\gamma}(t)  \nonumber \\
& = & \frac{\partial}{\partial t}|\Psi(t)\rangle - \frac{\partial}{\partial t}|{\rm D}^M_{2}(t)\rangle,
\label{deviation_1}
\end{eqnarray}
where the vectors $\vec{\chi}(t)$ and $\vec{\gamma}(t)$ obey the Schr\"{o}dinger equation  $\vec{\chi}(t)=\partial |\Psi(t)\rangle / \partial t = \frac{1}{i\hbar}\hat{H}|\Psi(t)\rangle$ and the Dirac-Frenkel variational dynamics $\vec{\gamma}(t)=\partial |{\rm D}^M_{2}\rangle / \partial t$, respectively.
The deviation vector $\vec{\delta}(t)$ can be calculated as
$
\vec{\delta}(t) = \frac{1}{i\hbar}\hat{H}|{\rm D}^M_{2}(t)\rangle - \frac{\partial}{\partial t}|{\rm D}^M_{2}(t)\rangle
$.
Thus, the accuracy of the trial state is indicated by the amplitude of the deviation vector $\Delta(t)=||\vec{\delta}(t)||$.
In order to view the deviation, a dimensionless relative deviation $\sigma$ is calculated as
\begin{equation}
\sigma = \frac{{\rm max}\{\Delta(t)\} }{{\rm mean}\{N_{\rm err}(t)\}}, \quad \quad t \in [0, t_{\rm max}].
\label{relative_error}
\end{equation}
where $N_{\rm err}(t)=||\vec{\chi}(t)||$ is the amplitude of the time derivative of the wave function,
\begin{eqnarray}
N_{\rm err}(t) & = & \sqrt{-\langle\frac{\partial}{\partial t}\Psi(t)|\frac{\partial}{\partial t}\Psi(t)\rangle} \nonumber \\
& = & \sqrt{\langle {\rm D}^M_{2}(t)|\hat{H}^2|{\rm D}^M_{2}(t)\rangle}.
\end{eqnarray}

\section{The multi-D$_2$ Davydov trial state}
\setcounter{equation}{0}

\begin{figure}[tbp]
\centering
\includegraphics[scale=0.47]{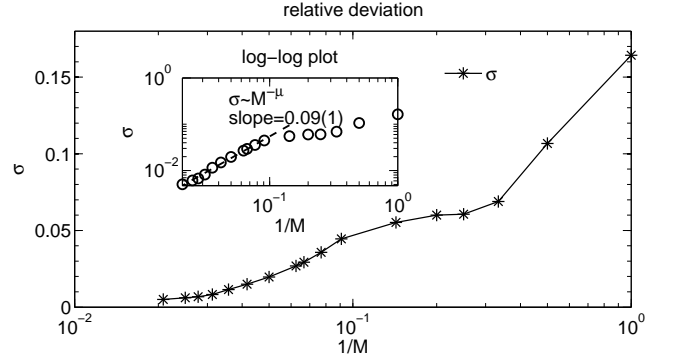}
\caption{The relative deviation $\sigma$ of the multi-${\rm D}_2$ {\it Ansatz} is displayed as a function of $1/M$. The parameters are set to be $\hbar\omega_{\rm diag}=90~{\rm meV}$, $\hbar\omega_{\rm vib2}=65~{\rm meV}$, $J_{\rm S_1,TT}=20~{\rm meV}$, $\epsilon_{\rm S_{1},g}-\epsilon_{\rm TT,g}=100~{\rm meV}$, $\gamma^{-1}_{\rm diag}=\gamma^{-1}_{\rm o.d.}=1~{\rm ps}$, $S_{\rm S_{1}}=0.7$, $S_{\rm TT}=1.4$ and $S_{\rm S_{1},TT}^{\rm off}=0.1$.}
\label{relativenormerror}
\end{figure}

\label{Validity of variational dynamics}
We first test the accuracy of our multi-D${_2}$ {\it Ansatz} with parameters for realistic models. The relative deviation $\sigma$ (Eq.~\ref{relative_error}) is used to quantify how faithfully our result follows the Schr\"odinger equation. As shown in Fig.~\ref{relativenormerror}, the largest relative deviation $\sigma$ is found for the single D$_2$ {\it Ansatz}, and the relative deviation $\sigma$ decreases and goes to zero as the multiplicity $M$ approaches infinity. The log-log plot of ($\sigma,1/M$) (inset) indicates a power-law relationship with an exponent of $\mu$, further inferring a numerically exact solution in the limit of $M \to \infty$.

\begin{figure}[tbp]
\centering
\includegraphics[scale=0.47]{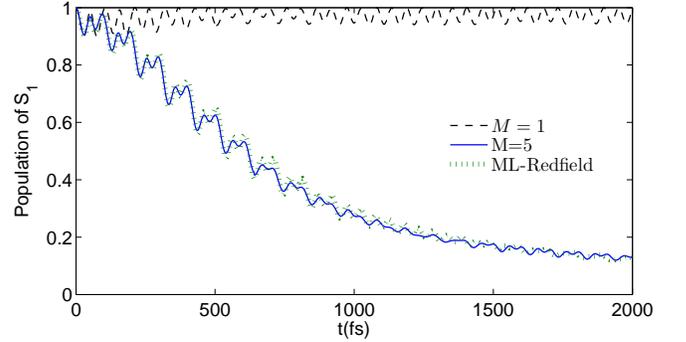}
\caption{Time evolution of singlet population calculated by the multilevel Red-field approach at $T=1K$ and the multi-$D_2$ {\it Ansatz} at $T=0K$. The frequency of the phonon mode is set to be $\hbar\omega_{\rm diag}=40~{\rm meV}$, $\hbar\omega_{\rm o.d.}=0~{\rm meV}$. The other parameters are fixed at $J_{\rm S_1,TT}=10~{\rm meV}$, $\epsilon_{\rm S_{1},g}-\epsilon_{\rm TT,g}=50~{\rm meV}$, $\gamma^{-1}_{\rm diag}=\gamma^{-1}_{\rm o.d.}=1~{\rm ps}$, $S_{\rm S_{1}}=0.7$, $S_{\rm TT}=1.4$.}
\label{compare_pop}
\end{figure}

In order to further support the accuracy of our results obtained by the variational approach, we compare the zero-temperature dynamics obtained by the multi-$D_2$ {\it Ansatz} with the multilevel Redfield results at $T=1K$. As show in Fig~\ref{compare_pop}, the population dynamics calculated by the $D^{M=5}_2$ {\it Ansatz} is in quantitative agreement with the multilevel Redfield results, yielding a much more accurate result than that with the single D$_2$ {\it Ansatz}. In addition, the relative deviation $\sigma$ in this case is smaller than $0.1$, small enough to guarantee the accuracy of the result by the multi-$D_2$ {\it Ansatz}. From the viewpoint of theoretical analysis, these calculated results satisfy the energy matching condition $\left(\nu-0.7\right)\hbar\omega_{\rm diag}=50~{\rm meV}$ in the case of $\nu=2$, which is obtained from the Fermi golden formula at the zero temperature limit and the slow vibrational relaxation limit, namely, $\gamma_{\rm diag}\rightarrow0$,
\begin{eqnarray}
k_{TT\leftarrow S_{1}}	=	\sum_{v=0}^{\infty}\frac{1}{\hbar^{2}}\left(J_{S_{1},TT}\left\langle \chi_{0}^{S_{1}}|\chi_{0}^{TT}\right\rangle \right)^{2}\nonumber\\	\times\delta\left(\epsilon_{S_{1},g}-\lambda_{S_{1},g}-\epsilon_{TT,g}+\lambda_{TT,g}-\nu\hbar\omega_{\rm diag}\right)
\end{eqnarray}
\label{fermi-golden}
where $\left|\left\langle \chi_{0}^{S_{1}}|\chi_{0}^{TT}\right\rangle \right|^{2}$ is the Frank-Condon factor associated with the vibronic transition from the $0$-th vibrational level on the $S_1$ to the $\nu$th vibrational level on $TT$.\cite{yuta_2017} Thus for the case shown in Fig.~\ref{compare_pop}, the transition from $S_1$ to $TT$ is driven by the phonon mode at around $\hbar\omega_{\rm diag}=40~{\rm meV}$ satisfying the condition in which the energy of $S_1$ including the vibrational energy matches its counterpart in the $TT$ state.

\begin{figure}[tbp]
\centering
\includegraphics[scale=0.47]{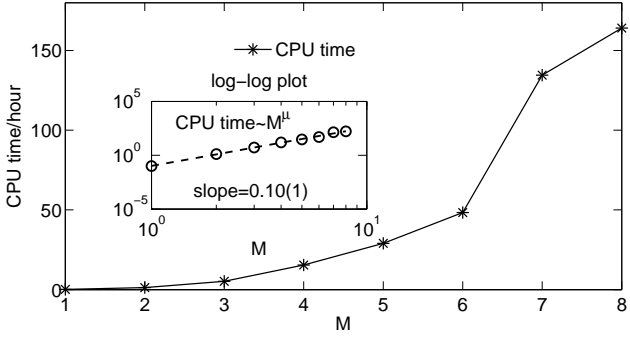}
\caption{CPU time of the multi-${\rm D}_2$ {\it Ansatz} is displayed as a function of $M$. Simulations of dynamical processes are on a long timescale of $2000~{\rm fs}$ and $90$ phonon modes are included in the calculations. In the inset, the relationship of CPU time$\sim M^{\mu}$ is displayed on a log-log scale and the dashed line represents a power-law fit.}
\label{cputime}
\end{figure}
The computational time and memory consumption for the variational method is also investigated to gain an understanding of the computational cost. As shown in Fig.\ref{cputime}, a power-law increase of the run time has been found for the variational approach against the multiplicity $M$ of the multi-D${_2}$ trial states. In the calculation, time constant of the vibrational dephasing is set to be $\gamma^{-1}_{\rm o.d.}=1~{\rm ps}$. The parameters are set to be $\hbar\omega_{\rm diag}=60~{\rm meV}$, $\hbar\omega_{\rm o.d.}=30~{\rm meV}$, $J_{\rm S_1,TT}=20~{\rm meV}$, $\epsilon_{\rm S_{1},g}-\epsilon_{\rm TT,g}=100~{\rm meV}$ and $S_{\rm S_{1},TT}^{\rm o.d.}=0.1$. Memory consumption can be estimated based on the number of variables in the calculation. There are nearly $(4M+2MN)^2+2(4M+2MN)+10MN+9M$ variables in our program, where each variable occupies $16$ bytes of the memory. For the number of phonon bath modes $N=200$ and the multiplicity of the D$_2$ trial states $M=12$, one has only $358$ MB (million bytes), which is still not too large for computing.

\begin{figure}[tbp]
\centering
\includegraphics[scale=0.47]{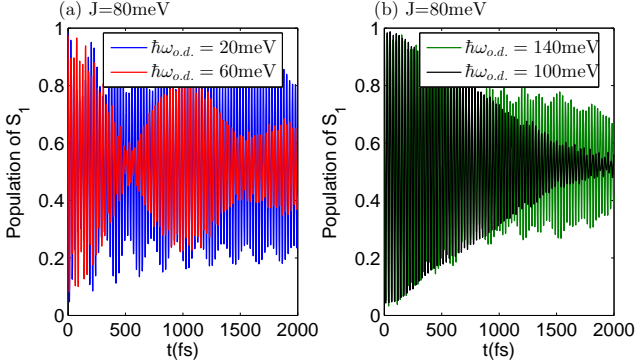}
\caption[FIG]{Time evolution of the singlet population for the case of strong excitonic coupling $J_{\rm S_1,TT}=80~{\rm meV}$ and $\epsilon_{\rm S_{1},g}-\epsilon_{\rm TT,g}=30~{\rm meV}$ with only off-diagonal coupling. The phonon modes $\hbar\omega_{\rm o.d.}$ are (a) $20~{\rm meV}$ and $60~{\rm meV}$, and (b) $100~{\rm meV}$ and $140~{\rm meV}$}
\label{population_largeJ_off}
\end{figure}

If excitonic coupling is strong, efficient SF is not found under the influence of off-diagonal exciton-phonon coupling, though both diagonal and off-diagonal coupling could aid efficient SF if excitonic coupling is weak. As shown in Fig.~\ref{population_largeJ_off}, when excitonic coupling is strong, the off-diagonal coupling is not able to trap the system in the $\rm TT$ state, and thus no efficient SF is obtained.

\section{Equations of motion}
\setcounter{equation}{0}

In order to apply the Dirac-Frenkel time-dependent variational principle, we first calculate the Lagrangian $L_2$,
\begin{eqnarray}
\label{Lagrangian_detail}
&&L_2=\frac{i}{2}\hbar\sum_{i,j}\left(A_{j}^{*}\dot{A}_{i}-\dot{A}_{j}^{\ast}A_{i}+B_{j}^{\ast}\dot{B}_{i}-\dot{B}_{j}^{\ast}B_{i}\right)S_{ji}\nonumber\\ &&+\frac{i}{2}\hbar\sum_{i,j}\left(A_{j}^{\ast}A_{i}+B_{j}^{\ast}B_{i}\right)\sum_{q}[\frac{\dot{f}_{jq}^{\ast}f_{jq}+f_{jq}^{\ast}\dot{f}_{jq}}{2}\nonumber\\
&&-\frac{\dot{f}_{iq}f_{iq}^{\ast}+f_{iq}\dot{f}_{iq}^{\ast}}{2}+f_{jq}^{\ast}\dot{f}_{iq}-f_{iq}\dot{f}_{jq}^{\ast}]S_{ji}\nonumber\\
&&-\left\langle D_{2}^{M}\left(t\right)\right|\hat{H}\left|D_{2}^{M}\left(t\right)\right\rangle
\end{eqnarray}
where the Debye-Waller factor is $S_{ji}=\exp{\sum_{q}\left\{-\left(\left|f_{jq}\right|^{2}+\left|f_{iq}\right|^{2}\right)/2+f_{jq}^{\ast}f_{iq}\right\}}$,
and the last term in Eq.~(\ref{Lagrangian_detail}) can be obtained as
\begin{eqnarray}
&&\left\langle D_{2}^{M}\left(t\right)\right|H\left|D_{2}^{M}\left(t\right)\right\rangle\nonumber\\
&&=\sum_{i,j}\left(\epsilon_{S_{1}}A_{j}^{\ast}A_{i}+\epsilon_{TT}B_{j}^{\ast}B_{i}\right)S_{ji}+J\sum_{i,j}\left(A_{j}^{\ast}B_{i}+B_{j}^{\ast}A_{i}\right)S_{ji}\nonumber\\
&&+\sum_{i,j}\left(A_{j}^{\ast}A_{i}+B_{j}^{\ast}B_{i}\right)\sum_{q}\hbar\omega_{q}f_{jq}^{\ast}f_{iq}S_{ji}\nonumber\\ &&+\sum_{q}\sum_{i,j}\left(g_{S_{1,}q}A_{j}^{\ast}A_{i}+g_{TT,q}B_{j}^{\ast}B_{i}\right)\hbar\omega_{q}\left(f_{iq}+f_{jq}^{\ast}\right)S_{ji}\nonumber\\
&&+\sum_{i,j}\left(A_{j}^{\ast}B_{i}+B_{j}^{\ast}A_{i}\right)\sum_{q}c_{q}\hbar\omega_{q}\left(f_{iq}+f_{jq}^{\ast}\right)S_{ji}
\end{eqnarray}

The Dirac-Frenkel variational principle results in equations of motion for $A_i$ and $B_i$,
\begin{eqnarray}
&&-i\hbar\sum_{i}\dot{A}{}_{i}S_{ki}\nonumber \\	&&-\frac{i}{2}\hbar\sum_{i}A_{i}\sum_{q}\left[-\left(\dot{f}_{iq}f_{iq}^{\ast}+f_{iq}\dot{f}_{iq}^{\ast}\right)+2f_{kq}^{\ast}\dot{f}_{iq}\right]S_{ki}\nonumber \\
&&=-\epsilon_{S_{1}}\sum_{i}A_{i}S_{ki}-J\sum_{i}B_{i}S_{ki}-\sum_{i}A_{i}\sum_{q}\hbar\omega_{q}f_{kq}^{\ast}f_{iq}S_{ki}\nonumber\\
&&-\sum_{i}A_{i}\sum_{q}g_{S_{1},q}\hbar\omega_{q}\left(f_{iq}+f_{kq}^{\ast}\right)S_{ki}\nonumber\\
&&-\sum_{i}B_{i}\sum_{q}c_{q}\hbar\omega_{q}\left(f_{iq}+f_{kq}^{\ast}\right)S_{ki}
\end{eqnarray}
and
\begin{eqnarray}
&&-i\hbar\sum_{i}\dot{B}_{i}S_{ki}\nonumber \\ &&-\frac{i}{2}\hbar\sum_{i}B_{i}\sum_{q}\left[-\left(\dot{f}_{iq}f_{iq}^{\ast}+f_{iq}\dot{f}_{iq}^{\ast}\right)+2f_{kq}^{\ast}\dot{f}_{iq}\right]S_{ki}\nonumber \\
&&=-\epsilon_{TT}\sum_{i}B_{i}S_{ki}-J\sum_{i}A_{i}S_{ki}-\sum_{i}B_{i}\sum_{q}\hbar\omega_{q}f_{kq}^{\ast}f_{iq}S_{ki}\nonumber\\ &&-\sum_{i}B_{i}\sum_{q}g_{TT,q}\hbar\omega_{q}\left(f_{iq}+f_{kq}^{\ast}\right)S_{ki}\nonumber\\
&&-\sum_{i}A_{i}\sum_{q}c_{q}\hbar\omega_{q}\left(f_{iq}+f_{kq}^{\ast}\right)S_{ki}
\end{eqnarray}
The equations of motion for $f_{iq}$ are
\begin{eqnarray} &&-i\hbar\sum_{i}\left(A_{k}^{\ast}\dot{A}_{i}+B_{k}^{\ast}\dot{B}_{i}\right)f_{iq}S_{ki}-i\hbar\sum_{i}\left(A_{k}^{\ast}A_{i}+B_{k}^{\ast}B_{i}\right)\dot{f}_{iq}S_{ki}\nonumber\\	&&-\frac{i}{2}\hbar\sum_{i}\left(A_{k}^{\ast}A_{i}+B_{k}^{\ast}B_{i}\right)f_{iq}S_{ki}\sum_{p}\left(2f_{kp}^{\ast}\dot{f}_{ip}-\dot{f}_{ip}f_{ip}^{\ast}-f_{ip}\dot{f}_{ip}^{\ast}\right)\nonumber\\
&&=-\sum_{i}\left(\epsilon_{S_{1}}A_{k}^{\ast}A_{i}+\epsilon_{TT}B_{k}^{\ast}B_{i}\right)f_{iq}S_{ki}\\
&&-J\sum_{i}\left(A_{k}^{\ast}B_{i}+B_{k}^{\ast}A_{i}\right)f_{iq}S_{ki}\nonumber\\	&&-\sum_{i}\left(g_{S_{1},q}A_{k}^{\ast}A_{i}+g_{TT,q}B_{k}^{\ast}B_{i}\right)\hbar\omega_{q}S_{ki}\nonumber\\	
&&-\sum_{i}f_{iq}\sum_{p}\left(g_{S_{1},p}A_{k}^{\ast}A_{i}+g_{TT,p}B_{k}^{\ast}B_{i}\right)\hbar\omega_{p}\left(f_{ip}+f_{kp}^{\ast}\right)S_{ki}\nonumber\\	&&-\sum_{i}\left(A_{k}^{\ast}B_{i}+B_{k}^{\ast}A_{i}\right)c_{q}\hbar\omega_{q}S_{ki}\nonumber\\
&&-\sum_{i}\left(A_{k}^{\ast}B_{i}+B_{k}^{\ast}A_{i}\right)f_{iq}\sum_{p}c_{p}\hbar\omega_{p}\left(f_{ip}+f_{kp}^{\ast}\right)S_{ki}
\end{eqnarray}
---


\begin{thebibliography}{999}
\bibitem{smith_2010}Smith, M. B.; Michl, J. Singlet Fission. Chem. Rev. \textbf{2010}, 110 (11), 6891-6936.



\bibitem{smith_2013}Smith, M. B.; Michl, J. Recent Advances in Singlet Fission. Annu. Rev. Phys. Chem. \textbf{2013}, 64, 361-86.


\bibitem{chan_2013}Chan, W.-L.; Berkelbach, T. C.; Provorse, M. R.; Monahan, N. R.; Tritsch, J. R.; Hybertsen, M. S.; Reichman, D. R.; Gao, J.; Zhu, X.-Y. The Quantum Coherent Mechanism for Singlet Fission: Experiment and Theory. Acc. Chem. Res. \textbf{2013}, 46 (6), 1321-1329.



\bibitem{singh_1965}Singh, S.; Jones, W. J.; Siebrand, W.; Stoicheff, B. P.; Schneider, W. G. Laser Generation of Excitons and Fluorescence in Anthracene Crystals. J. Chem. Phys. \textbf{1965}, 42 (1), 330-342.



\bibitem{hanna_2006}Hanna, M. C.; Nozik, A. J. Solar Conversion Efficiency of Photovoltaic and Photoelectrolysis Cells with Carrier Multiplication Absorbers. J. Appl. Phys. \textbf{2006}, 100 (7), 74510.



\bibitem{congreve2013external}Congreve, D. N.; Lee, J.; Thompson, N. J.; Hontz, E.; Yost, S. R.; Reusswig, P. D.; Bahlke, M. E.; Reineke, S.; Van Voorhis, T.; Baldo, M. A. External Quantum Efficiency Above 100\% in a Singlet-Exciton-Fission-Based Organic Photovoltaic Cell. Science \textbf{2013}, 340 (6130), 334-337.




\bibitem{piland_2014}Piland, G. B.; Burdett, J. J.; Dillon, R. J.; Bardeen, C. J. Singlet Fission: From Coherences to Kinetics. J. Phys. Chem. Lett. \textbf{2014}, 5 (13), 2312-2319.



\bibitem{ramanan_2011}Ramanan, C.; Smeigh, A. L.; Anthony, J. E.; Marks, T. J.; Wasielewski, M. R. Competition between Singlet Fission and Charge Separation in Solution-Processed Blend Films of 6,13-Bis(triisopropylsilylethynyl)pentacene with Sterically-Encumbered Perylene-3,4:9,10-Bis(dicarboximide)s. J. Am. Chem. Soc. \textbf{2012}, 134 (1), 386-397.



\bibitem{billon_2013}Dillon, R. J.; Piland, G. B.; Bardeen, C. J. Different Rates of Singlet Fission in Monoclinic versus Orthorhombic Crystal Forms of Diphenylhexatriene. J. Am. Chem. Soc. \textbf{2013}, 135 (46), 17278-17281.



\bibitem{eaton_2013}Eaton, S. W.; Shoer, L. E.; Karlen, S. D.; Dyar, S. M.; Margulies, E. A.; Veldkamp, B. S.; Ramanan, C.; Hartzler, D. A.; Savikhin, S.; Marks, T. J.; et al. Singlet Exciton Fission in Polycrystalline Thin Films of a Slip-Stacked Perylenediimide. J. Am. Chem. Soc.
    \textbf{2013}, 135 (39), 14701-14712.




\bibitem{eaton_2015}Eaton, S. W.; Miller, S. A.; Margulies, E. A.; Shoer, L. E.; Schaller, R. D.; Wasielewski, M. R. Singlet Exciton Fission in Thin Films of Tert-Butyl-Substituted Terrylenes. J. Phys. Chem. A \textbf{2015}, 119 (18), 4151-4161.




\bibitem{wilson2011ultrafast}Wilson, M. W. B.; Rao, A.; Clark, J.; Kumar, R. S. S.; Brida, D.; Cerullo, G.; Friend, R. H. Ultrafast Dynamics of Exciton Fission in Polycrystalline Pentacene. J. Am. Chem. Soc. \textbf{2011}, 133 (31), 11830-11833.



\bibitem{chan2011observing}Chan, W.-L.; Ligges, M.; Jailaubekov, A.; Kaake, L.; Miaja-Avila, L.; Zhu, X.-Y. Observing the Multiexciton State in Singlet Fission and Ensuing Ultrafast Multielectron Transfer. Science \textbf{2011}, 334 (6062), 1541-1545.



\bibitem{chan2012energy}Chan, W.-L.; Ligges, M.; ZhuX-Y. The Energy Barrier in Singlet Fission Can Be Overcome through Coherent Coupling and Entropic Gain. Nat Chem \textbf{2012}, 4 (10), 840-845.



\bibitem{wilson2013temperature}Wilson, M. W. B.; Rao, A.; Johnson, K.; G¨¦linas, S.; di Pietro, R.; Clark, J.; Friend, R. H. Temperature-Independent Singlet Exciton Fission in Tetracene. J. Am. Chem. Soc. \textbf{2013}, 135 (44), 16680-16688.



\bibitem{ma2012rubrene}Ma, L.; Zhang, K.; Kloc, C.; Sun, H.; Michel-Beyerle, M. E.; Gurzadyan, G. G. Singlet Fission in Rubrene Single Crystal: Direct Observation by Femtosecond Pump-Probe Spectroscopy. Phys. Chem. Chem. Phys. \textbf{2012}, 14 (23), 8307-8312.



\bibitem{greyson_2010}Greyson, E. C.; Vura-Weis, J.; Michl, J.; Ratner, M. A. Maximizing Singlet Fission in Organic Dimers: Theoretical Investigation of Triplet Yield in the Regime of Localized Excitation and Fast Coherent Electron Transfer. J. Phys. Chem. B \textbf{2010}, 114 (45), 14168-14177.



\bibitem{teichen2012microscopic}Teichen, P. E.; Eaves, J. D. A Microscopic Model of Singlet Fission. J. Phys. Chem. B \textbf{2012}, 116 (37), 11473-11481.



\bibitem{berkelbach_2013_1}Berkelbach, T. C.; Hybertsen, M. S.; Reichman, D. R. Microscopic Theory of Singlet Exciton Fission. I. General Formulation. J. Chem. Phys. \textbf{2013}, 138 (11), 114102.



\bibitem{berkelbach_2013_2}Berkelbach, T. C.; Hybertsen, M. S.; Reichman, D. R. Microscopic Theory of Singlet Exciton Fission. II. Application to Pentacene Dimers and the Role of Superexchange. J. Chem. Phys. \textbf{2013}, 138 (11), 114103.




\bibitem{berkelbach2014microscopic3}Berkelbach, T. C.; Hybertsen, M. S.; Reichman, D. R. Microscopic Theory of Singlet Exciton Fission. III. Crystalline Pentacene. J. Chem. Phys. \textbf{2014}, 141 (7), 74705.



\bibitem{mirjani2014theoretical}Mirjani, F.; Renaud, N.; Gorczak, N.; Grozema, F. C. Theoretical Investigation of Singlet Fission in Molecular Dimers: The Role of Charge Transfer States and Quantum Interference. J. Phys. Chem. C \textbf{2014}, 118 (26), 14192-14199.



\bibitem{tao2014electronically}Tao, G. Electronically Nonadiabatic Dynamics in Singlet Fission: A Quasi-Classical Trajectory Simulation. J. Phys. Chem. C \textbf{2014}, 118 (31), 17299-17305.



\bibitem{tamura2015first}Tamura, H.; Huix-Rotllant, M.; Burghardt, I.; Olivier, Y.; Beljonne, D. First-Principles Quantum Dynamics of Singlet Fission: Coherent versus Thermally Activated Mechanisms Governed by Molecular \ensuremath{\pi} Stacking. Phys. Rev. Lett. \textbf{2015}, 115 (10), 107401.



\bibitem{fujihashi2016fluctuations}Fujihashi, Y.; Ishizaki, A. Fluctuations in Electronic Energy Affecting Singlet Fission Dynamics and Mixing with Charge-Transfer State: Quantum Dynamics Study. J. Phys. Chem. Lett. \textbf{2016}, 7 (3), 363-369.



\bibitem{yao2016coherent}Yao, Y. Coherent Dynamics of Singlet Fission Controlled by Nonlocal Electron-Phonon Coupling. Phys. Rev. B \textbf{2016}, 93 (11), 115426.



\bibitem{zimmerman2010singlet}Zimmerman, P. M.; Zhang, Z.; Musgrave, C. B. Singlet Fission in Pentacene through Multi-Exciton Quantum States. Nat Chem \textbf{2010}, 2 (8), 648-652.



\bibitem{zimmerman_2011}Zimmerman, P. M.; Bell, F.; Casanova, D.; Head-Gordon, M. Mechanism for Singlet Fission in Pentacene and Tetracene: From Single Exciton to Two Triplets. J. Am. Chem. Soc. \textbf{2011}, 133 (49), 19944-19952.



\bibitem{feng2013fission}Feng, X.; Luzanov, A. V; Krylov, A. I. Fission of Entangled Spins: An Electronic Structure Perspective. J. Phys. Chem. Lett. \textbf{2013}, 4 (22), 3845-3852.



\bibitem{yost2014transferable}Yost, S. R.; Lee, J.; Wilson, M. W. B.; Wu, T.; McMahon, D. P.; Parkhurst, R. R.; Thompson, N. J.; Congreve, D. N.; Rao, A.; Johnson, K.; et al. A Transferable Model for Singlet-Fission Kinetics. Nat Chem \textbf{2014}, 6 (6), 492-497.



\bibitem{casanova2014electronic}Casanova, D. Electronic Structure Study of Singlet Fission in Tetracene Derivatives. J. Chem. Theory Comput. 2014, 10 (1), 324-334.



\bibitem{renaud2013mapping}Renaud, N.; Sherratt, P. A.; Ratner, M. A. Mapping the Relation between Stacking Geometries and Singlet Fission Yield in a Class of Organic Crystals. J. Phys. Chem. Lett. \textbf{2013}, 4 (7), 1065-1069.



\bibitem{wang2014maximizing}Wang, L.; Olivier, Y.; Prezhdo, O. V; Beljonne, D. Maximizing Singlet Fission by Intermolecular Packing. J. Phys. Chem. Lett. \textbf{2014}, 5 (19), 3345-3353.



\bibitem{lukman_2016_sidegroup}Lukman, S.; Chen, K.; Hodgkiss, J. M.; Turban, D. H. P.; Hine, N. D. M.; Dong, S.; Wu, J.; Greenham, N. C.; Musser, A. J. Tuning the Role of Charge-Transfer States in Intramolecular Singlet Exciton Fission through Side-Group Engineering. Nat. Commun. \textbf{2016}, 7, 13622.



\bibitem{monahan2015charge}Monahan, N.; Zhu, X.-Y. Charge Transfer-Mediated Singlet Fission. Annu. Rev. Phys. Chem. \textbf{2015}, 66 (1), 601-618.



\bibitem{busby2015intraSF}Busby, E.; Xia, J.; Wu, Q.; Low, J. Z.; Song, R.; Miller, J. R.; Zhu, X.-Y.; Campos, L. M.; Sfeir, M. Y. A Design Strategy for Intramolecular Singlet Fission Mediated by Charge-Transfer States in Donor¨Cacceptor Organic Materials. Nat Mater \textbf{2015}, 14 (4), 426-433.



\bibitem{ethan_2014}Alguire, E. C.; Subotnik, J. E.; Damrauer, N. H. Exploring Non-Condon Effects in a Covalent Tetracene Dimer: How Important Are Vibrations in Determining the Electronic Coupling for Singlet Fission? J. Phys. Chem. A \textbf{2015}, 119 (2), 299-311.



\bibitem{sanders2016intraSF}Sanders, S. N.; Kumarasamy, E.; Pun, A. B.; Appavoo, K.; Steigerwald, M. L.; Campos, L. M.; Sfeir, M. Y. Exciton Correlations in Intramolecular Singlet Fission. J. Am. Chem. Soc. \textbf{2016}, 138 (23), 7289-7297.



\bibitem{varnavski2015intraSF}Varnavski, O.; Abeyasinghe, N.; Arag\'o, J.; Serrano-P\'erez, J. J.; Ort\'i, E.; L\'opez Navarrete, J. T.; Takimiya, K.; Casanova, D.; Casado, J.; Goodson, T. High Yield Ultrafast Intramolecular Singlet Exciton Fission in a Quinoidal Bithiophene. J. Phys. Chem. Lett. \textbf{2015}, 6 (8), 1375-1384.



\bibitem{tao_2016}Zeng, T. Through-Linker Intramolecular Singlet Fission: General Mechanism and Designing Small Chromophores. J. Phys. Chem. Lett. \textbf{2016}, 7 (21), 4405-4412.



\bibitem{tao_2016_1}Zeng, T.; Goel, P. Design of Small Intramolecular Singlet Fission Chromophores: An Azaborine Candidate and General Small Size Effects. J. Phys. Chem. Lett. \textbf{2016}, 7 (7), 1351-1358.



\bibitem{eric_2016}Fuemmeler, E. G.; Sanders, S. N.; Pun, A. B.; Kumarasamy, E.; Zeng, T.; Miyata, K.; Steigerwald, M. L.; Zhu, X.-Y.; Sfeir, M. Y.; Campos, L. M.; et al. A Direct Mechanism of Ultrafast Intramolecular Singlet Fission in Pentacene Dimers. ACS Cent. Sci. \textbf{2016}, 2 (5), 316-324.



\bibitem{samuel2016intraSF}Ito, S.; Nagami, T.; Nakano, M. Design Principles of Electronic Couplings for Intramolecular Singlet Fission in Covalently-Linked Systems. J. Phys. Chem. A \textbf{2016}, 120 (31), 6236-6241.



\bibitem{meisner2012dpdc}Meisner, J. S.; Sedbrook, D. F.; Krikorian, M.; Chen, J.; Sattler, A.; Carnes, M. E.; Murray, C. B.; Steigerwald, M.; Nuckolls, C. Functionalizing Molecular Wires: A Tunable Class of $\alpha${,}$\omega$-Diphenyl-$\mu${,}$\nu$-Dicyano-Oligoenes. Chem. Sci. \textbf{2012}, 3 (4), 1007-1014.



\bibitem{tuan_2014_crystaleffect}Trinh, M. T.; Zhong, Y.; Chen, Q.; Schiros, T.; Jockusch, S.; Sfeir, M. Y.; Steigerwald, M.; Nuckolls, C.; Zhu, X. Intra- to Intermolecular Singlet Fission. J. Phys. Chem. C \textbf{2015}, 119 (3), 1312-1319.



\bibitem{yuta_2017}Fujihashi, Y.; Chen, L.; Ishizaki, A.; Wang, J.; Zhao, Y. Effect of High-Frequency Modes on Singlet Fission Dynamics. J. Chem. Phys. \textbf{2017}, 146 (4), 44101.



\bibitem{musser_2015}Musser, A. J.; Liebel, M.; Schnedermann, C.; Wende, T.; Kehoe, T. B.; Rao, A.; Kukura, P. Evidence for Conical Intersection Dynamics Mediating Ultrafast Singlet Exciton Fission. Nat Phys \textbf{2015}, 11 (4), 352-357.



\bibitem{bakulin2016real}Bakulin, A. A.; Morgan, S. E.; Kehoe, T. B.; Wilson, M. W. B.; Chin, A. W.; Zigmantas, D.; Egorova, D.; Rao, A. Real-Time Observation of Multiexcitonic States in Ultrafast Singlet Fission Using Coherent 2D Electronic Spectroscopy. Nat Chem \textbf{2016}, 8 (1), 16-23.



\bibitem{monahan2017dynamics}Monahan, N. R.; Sun, D.; Tamura, H.; Williams, K. W.; Xu, B.; Zhong, Y.; Kumar, B.; Nuckolls, C.; Harutyunyan, A. R.; Chen, G.; et al. Dynamics of the Triplet-Pair State Reveals the Likely Coexistence of Coherent and Incoherent Singlet Fission in Crystalline Hexacene. Nat Chem \textbf{2017}, 9 (4), 341-346.




\bibitem{tempelaar2017vibronic1}Tempelaar, R.; Reichman, D. R. Vibronic Exciton Theory of Singlet Fission. I. Linear Absorption and the Anatomy of the Correlated Triplet Pair State. J. Chem. Phys. \textbf{2017}, 146 (17), 174703.



\bibitem{tempelaar2017vibronic2}Tempelaar, R.; Reichman, D. R. Vibronic Exciton Theory of Singlet Fission. II. Two-Dimensional Spectroscopic Detection of the Correlated Triplet Pair State. J. Chem. Phys. \textbf{2017}, 146 (17), 174704.




\bibitem{morrison_2017}Morrison, A. F.; Herbert, J. M. Evidence for Singlet Fission Driven by Vibronic Coherence in Crystalline Tetracene. J. Phys. Chem. Lett. \textbf{2017}, 8, 1442-1448.



\bibitem{elenewski2017functional}Elenewski, J. E.; Cubeta, U. S.; Ko, E.; Chen, H. Functional Mode Singlet Fission Theory. J. Phys. Chem. C \textbf{2017}, 121 (8), 4130-4138.



\bibitem{renaud_2014}Renaud, N.; Grozema, F. C. Intermolecular Vibrational Modes Speed Up Singlet Fission in Perylenediimide Crystals. J. Phys. Chem. Lett. \textbf{2015}, 6 (3), 360-365.



\bibitem{mu_85}Munn, R. W.; Silbey, R. Theory of Electronic Transport in Molecular Crystals. II. Zeroth Order States Incorporating Nonlocal Linear Electron-phonon Coupling. J. Chem. Phys. \textbf{1985}, 83 (4), 1843.



\bibitem{zh_12}Zhao, Y.; Luo B.; Zhang Y.; Ye J. Dynamics of a Holstein polaron with off-diagonal coupling, J. Chem. Phys. \textbf{2012}, 137, 084113.



\bibitem{zh_97}(a) Zhao, Y.; Brown, D. W.; Lindenberg, K. Variational Energy Band Theory for Polarons : Mapping Polaron Structure with the Toyozawa Method Variational Energy Band Theory for Polarons : Mapping Polaron Structure with the Toyozawa Method. J. Chem. Phys. \textbf{1997}, 107 (8), 3159; (b) Brown, D. W.; Lindenberg, K.; Zhao, Y. Variational Energy Band Theory for Polarons : Mapping Polaron Structure with the Global-Local Method Variational Energy Band Theory for Polarons : Mapping Polaron Structure with the Global-Local Method. J. Chem. Phys. \textbf{1997}, 107 (8), 3179.



\bibitem{zhou2015polaron}Zhou, N;. Huang, Z.; Zhu, J.; Chernyak, V.; Zhao, Y. Polaron dynamics with a multitude of Davydov D$_2$ trial states, J. Chem. Phys. \textbf{2015}, 143, 014113.




\bibitem{zhou2016fast}Zhou, N; Chen, L.; Huang, Z.; Sun, K.;  Tanimura, Y.; Zhao, Y. Fast, Accurate Simulation of Polaron Dynamics and Multidimensional Spectroscopy by Multiple Davydov Trial States, J. Phys. Chem. A, \textbf{2016}, 120 1562-1576.



\bibitem{bera2014stabilizing}Bera, S.; Florens, S.; Baranger, H. U.; Roch, N.; Nazir, A.; Chin, A. W. Stabilizing Spin Coherence through Environmental Entanglement in Strongly Dissipative Quantum Systems. Phys. Rev. B \textbf{2014}, 89 (12), 121108(R).



\bibitem{zhou2014ground}Zhou, N.; Chen, L.; Zhao, Y.; Mozyrsky, D.; Chernyak, V.; Zhao, Y. Ground-State Properties of Sub-Ohmic Spin-Boson Model with Simultaneous Diagonal and off-Diagonal Coupling. Phys. Rev. B \textbf{2014}, 90 (15), 155135.



\bibitem{wang2016variational}Wang, L.; Fujihashi, Y.; Chen, L.; Zhao, Y. Finite-Temperature Time-Dependent Variation with Multiple Davydov States. J. Chem. Phys. \textbf{2017}, 146 (12), 124127.



\bibitem{huang2017polaron}Huang, Z.; Wang, L.; Wu, C.-Q.; Chen, L.; Grossmann, F.; Zhao, Y. Polaron Dynamics with off-Diagonal Coupling: Beyond the Ehrenfest Approximation. Phys. Chem. Chem. Phys. \textbf{2016}, 19 (2), 1655-1668.



\bibitem{huang2017}Huang, Z.; Chen, L.; Zhou, N.; Zhao, Y. Transient Dynamics of a One-Dimensional Holstein Polaron under the Influence of an External Electric Field. Ann. Phys. \textbf{2017}, 529 (3), 1600367.



\bibitem{yamagata2011}Yamagata, H.; Norton, J.; Hontz, E.; Olivier, Y.; Beljonne, D.; Br\'edas, J. L.; Silbey, R. J.; Spano, F. C. The Nature of Singlet Excitons in Oligoacene Molecular Crystals. J. Chem. Phys. \textbf{2011}, 134 (20), 204703.



\bibitem{beljonne2013}Beljonne, D.; Yamagata, H.; Br\'edas, J. L.; Spano, F. C.; Olivier, Y. Charge-Transfer Excitations Steer the Davydov Splitting and Mediate Singlet Exciton Fission in Pentacene. Phys. Rev. Lett. \textbf{2013}, 110 (22), 226402.



\bibitem{chen2016optical}Chen, L.; Lu, J.; Long, G.; Zheng, F.; Zhang, J.; Zhao, Y. Optical and Transport Properties of Single Crystal Rubrene: A Theoretical Study. Chem. Phys. \textbf{2016}, 481, 198-205.



\bibitem{ito_2015}Ito, S.; Nagami, T.; Nakano, M. Density Analysis of Intra- and Intermolecular Vibronic Couplings toward Bath Engineering for Singlet Fission. J. Phys. Chem. Lett. \textbf{2015}, 6 (24), 4972-4977.



\bibitem{vallet2013tunable}Vallett, P. J.; Snyder, J. L.; Damrauer, N. H. Tunable Electronic Coupling and Driving Force in Structurally Well-Defined Tetracene Dimers for Molecular Singlet Fission: A Computational Exploration Using Density Functional Theory. J. Phys. Chem. A \textbf{2013}, 117 (42), 10824-10838.



\bibitem{tamura2011chargetransfer}Tamura, H.; Burghardt, I.; Tsukada, M. Exciton Dissociation at Thiophene/Fullerene Interfaces: The Electronic Structures and Quantum Dynamics. J. Phys. Chem. C \textbf{2011}, 115 (20), 10205-10210.




\bibitem{tamura2012chargetransfer}Tamura, H.; Martinazzo, R.; Ruckenbauer, M.; Burghardt, I. Quantum Dynamics of Ultrafast Charge Transfer at an Oligothiophene-Fullerene Heterojunction. J. Chem. Phys. \textbf{2012}, 137 (22), 22A540.

\end{thebibliography}
\end{document}